\documentstyle[12pt]{article}
\newlength{\mytopmargin}
\newlength{\myleftmargin}
\def\zz{\rlx\hbox{\small \sf Z\kern-.4em Z}}

\begin{document}

\vspace{1cm}
\noindent
\begin{center}{   \large \bf
Random matrix ensembles with an effective extensive \\ external charge}  
\end{center}
\vspace{5mm}

\noindent
{\center T.H.~Baker,
 P.J.~Forrester
 and P.A.~Pearce
\\
\it Department of Mathematics, University of Melbourne, Parkville, Victoria
3052, Australia}
\vspace{.5cm}

\small
\begin{quote}
Recent theoretical studies of chaotic scattering have encounted ensembles
of random matrices in which the eigenvalue probability density function
contains a one-body factor with an exponent proportional to the number of
eigenvalues. Two such ensembles have been encounted: an ensemble of
unitary matrices specified by the so-called Poisson kernel, and the
Laguerre ensemble of positive definite matrices. Here we consider various
properties of these ensembles. Jack polynomial theory is used to prove
a reproducing property of the Poisson kernel, and a certain unimodular
mapping is used to demonstrate that the variance of a linear statistic is
the same as in the Dyson circular ensemble. For the Laguerre ensemble, the
scaled global density is calculated exactly for all even values of the
parameter $\beta$, while for $\beta = 2$ (random matrices with unitary
symmetry), the neighbourhood of the smallest eigenvalue is shown to
be in the soft edge universality class.
\end{quote}

\vspace{.5cm}
\noindent
\section{Introduction}
\setcounter{equation}{0}
In the theory of random matrices, a primary task is to compute the
probability density function (p.d.f.) for the eigenvalues from knowledge
of the p.d.f.~for the ensemble of matrices. Two examples of random matrix
ensembles of interest in this paper are Dyson's \cite{Dy62} circular
ensembles of symmetric, unitary and self dual quaternion unitary random
matrices (labelled by $\beta=1$, 2 and 4 respectively), and the Laguerre
ensemble of random Wishart matrices $A = X^\dagger X$, where $X$ is a
random $M \times N$ ($M \ge N$) matrix which has either real
($\beta = 1$), complex ($\beta = 2$) or quaternion real ($\beta = 4$)
Gaussian random elements. In the circular ensemble, the p.d.f.~for the
matrices is uniquely specified by requiring that it be uniform and
unchanged by mappings of the form $U \mapsto V U V'$, where $V$ is an
arbitrary $N \times N$ unitary matrix and $V' = V^T$ for $\beta = 1$,
$V'$ is arbitrary for $\beta = 2$ and
$V' = V^D$ for $\beta = 4$ ($D$ denotes the quaternion dual). In the Laguerre
ensemble, the distribution of the elements of $X$ are taken to be the
Gaussian 
$
A e^{-\beta {\rm Tr}(X^\dagger X)/2}
$
which is equivalent to choosing each element independently with a
Gaussian distribution $A' e^{-\beta |x_{jk}|^2/2}$.

The corresponding p.d.f.~for the eigenvalues $e^{i \theta_j}$, $j=1,\dots,N$
in the circular ensemble is
\begin{equation}\label{1}
{1 \over C_{\beta N}} \prod_{1 \le j < k \le N} | e^{i \theta_k} -
e^{i \theta_j}|^\beta,
\end{equation}
while for the Laguerre ensemble the p.d.f.~is given by
\begin{equation}\label{2}
{1 \over C_{a \beta N}'}
\prod_{j=1}^N e^{-\beta \lambda_j / 2} \lambda_j^{a \beta /2}
\prod_{1 \le j < k \le N} |\lambda_k - \lambda_j|^\beta, \quad
\lambda_j > 0,
\end{equation}
where $a := M - N +1 -2/\beta$. In both cases the eigenvalue p.d.f.'s can be
written in the form of a Boltzmann factor for a classical gas, with potential
energy consisting of one and two body terms only and interacting at inverse
temperature $\beta$,
\begin{equation}\label{2.1}
\exp \bigg ( - \beta \Big ( \sum_{j=1}^N V_1(x_j) + \sum_{1 \le j < k \le N}
V_2(x_j,x_k) \Big ) \bigg ).
\end{equation}
Thus for the circular ensemble
\begin{equation}\label{3}
 V_1(x) = 0, \quad V_2(x,y) = - \log |e^{ix} - e^{iy} |,
\end{equation}
while for the Laguerre ensemble
\begin{equation}\label{4}
V_1(x) = {x \over 2} - {a \over 2} \log x, \quad V_2(x,y) = - \log |x-y|,
\end{equation}
which displays the well known fact that the analogous classical gas has a
two body logarithmic potential. The logarithmic potential is special in that
it is the Coulomb potential between like charges in two dimensions. In
(\ref{3}) the two dimensional charges are confined to a unit circle, while
in (\ref{4}) the two dimensional charges are free to move on the half line
$x > 0$, but are confined to the neighbourhood of the origin by the
one body potential.

In this work we will focus attention on a subclass of random matrix
ensembles with eigenvalue p.d.f.'s of the form (\ref{2.1}) in which the
one body potential contains a term proportional to 
$N \log | 1 - \mu^* e^{ix} |$ (unitary matrices) or
$-N \log |x|$ (Laguerre ensemble). Again the two body term is logarithmic.
In the classical gas these one body potentials can be interpreted as being
due to an external fixed charge at the point $1/\mu^*$ in the complex plane
with strength proportional to $-N$ (unitary matrices), and as an external charge fixed at the
origin of stength proportional to $N$ (Laguerre ensemble). We see from
(\ref{2}) that an example of a random matrix of this type in the Laguerre case
is a Wishart matrix with $X$ a rectangular matrix in which the number of
columns is some fixed fraction of the number of rows. In the theory of
random unitary matrices, this type of eigenvalue p.d.f.~results as
a special case of the the ensemble of random matrices defined by the Poisson
kernel
\begin{equation}\label{5}
{1 \over C} {\det(1 - \bar {S} \bar{S}^\dagger)^{\beta (N-1)/2 + 1}
\over | \det (1 -  \bar{S}^\dagger S) |^{\beta (N - 1) + 2} },
\end{equation}
where the notation $\bar {S}$ denotes the average of $S$. Random
unitary matrices with this p.d.f.~occur in the study of scattering
problems in nuclear physics \cite{MPS85} and mesoscopic systems \cite{Br95}.
In the case $\bar {S} = 0$, this reduces to the p.d.f.~specifying Dyson's
circular ensemble (all members equally probable). In the case that
$\bar {S} = \mu 1_N$, $|\mu| <1$ ($1_N$ denotes the
$N \times N$ unit matrix), the corresponding eigenvalue p.d.f.~is given by
\begin{equation}\label{6}
{1 \over C_{\beta N} }\prod_{j=1}^N {(1 - |\mu|^2)^{\beta a'/2} \over
|1 - \mu^* e^{i \theta} |^{\beta a'}}
 \prod_{1 \le j < k \le N} | e^{i \theta_k} -
e^{i \theta_j}|^\beta
\end{equation}
where $a' = (N - 1 + 2/\beta)$ and $C_{\beta N}$ is
as in (\ref{1}), and so in the classical gas picture
there is a fixed charge of opposite sign at the point $1/\mu^*$, and the 
magnitude of this charge is indeed proportional to $N$.

In Section 2 we will consider the eigenvalue p.d.f.~(\ref{6}) for the
Poisson kernel. First the special reproducing property of the p.d.f.'s
(\ref{5}) and (\ref{6}) for $\beta = 1,2$ and 4 will be revised, and 
extended to all $\beta >0$ in the case of (\ref{6}). Our tool here is
Jack polynomial theory \cite{Ma95}. Then we will consider the effect
of the $N$-dependent exponent in (\ref{6}) on the one and two body correlation
functions, as well as on the fluctuation formula for the variance of a
linear statistic.

In Section 3 a physical problem giving rise to the Laguerre ensemble in
which $a$ in (\ref{2}) is equal to $N$ will be revised. Then we will
revise known theorems in the cases $\beta = 1$ and $2$ for the global
density and the distribution of the largest and smallest
eigenvalues. Next known integral formulas for the density \cite{BF97}
at general even $\beta$,
deduced from the theory of generalized Selberg integrals \cite{Ka93} and
their relationship to Jack polynomial theory, will be analyzed in the
appropriate limit to deduce that the formula for the global density holds
independent of $\beta$. For the special coupling $\beta = 2$ the local
distribution functions
 in the neigbourhood of the smallest eigenvalue are analyzed and
shown to belong to the universality class of the soft edge, giving 
rise to the Airy kernel \cite{Fo93a, TW94a, KF97a}. 
We conclude the section with an analysis of
some nonlinear equations \cite{TW94a,TW94b},
which explicitly demonstrates the universality of the distribution of the
smallest eigenvalue.

\section{The Poisson kernel}
\setcounter{equation}{0}
\subsection{Physical origin of the Poisson kernel}
The scattering matrix for $N$ channels entering and leaving a chaotic
cavity via a non-ideal lead containing a tunnel barrier has as its
p.d.f.~the Poisson kernel (\ref{5}) \cite{Br95}. Also, in scattering
problems in nuclear physics, (\ref{5}) has been used \cite{MPS85} to describe
situations in which the average of the scattering matrix is non-zero.
It was in the latter problem that the Poisson kernel first appeared
in an application of random matrix theory. In \cite{MPS85} a requirement
for the p.d.f.~of the ensemble of scattering matrices was the
special reproducing property
\begin{equation}\label{2.1x}
f(\bar{S}) := f( \langle S \rangle) = \langle f(S) \rangle
\end{equation}
for $f$ analytic in $S$. It was noted that a result of Hua \cite{Hu63}
gives that
\begin{equation}\label{2.2}
f(\bar{S}) = 
{1 \over C} \int f(S) {\det(1_N - \bar {S} \bar{S}^\dagger)^{\beta (N-1)/2 + 1}
\over | \det (1_N -  \bar{S}^\dagger S) |^{\beta (N - 1) + 2} }
\mu(dS)
\end{equation}
where $\mu(dS)$ is the invariant measure associated with the
Dyson circular ensemble, and thus that the Poisson kernel 
exhibits the reproducing property (\ref{2.1x}).

\subsection{The reproducing property for general $\beta$}
To make any sense out of (\ref{2.2}) it is necessary that $\beta = 1,2$ or
4, so that the measure $\mu(dS)$ has meaning. However, in the
case $\bar{S} = \mu 1_N$, the corresponding eigenvalue p.d.f.~(\ref{6})
can be interpreted as a Boltzmann factor and it makes sense to consider all
$\beta >0$. In this case, for $\beta = 1,2$ and 4, (\ref{6}) used in (\ref{2.2})
gives that the reproducing property restricted to analytic functions of the 
eigenvalues $e^{i \theta_j}$ reads
\begin{equation}\label{2.3}
f(\mu,\dots,\mu) = {1 \over C_{\beta N}} 
\prod_{l=1}^N \int_0^{2 \pi} d \theta_l
{(1 - |\mu|^2)^{\beta a'/2} \over
|1 - \mu^* e^{i \theta_l} |^{\beta a'}} f(e^{i\theta_1},\dots,e^{i\theta_N})
 \prod_{1 \le j < k \le N} | e^{i \theta_k} -
e^{i \theta_j}|^\beta.
\end{equation}
This equation is well defined for all $\beta > 0$, but its validity has only
been established for $\beta = 1,2$ and 4. Here we will establish its
validity for general $\beta >0$, using properties of the orthogonal
polynomials associated with the p.d.f.~(\ref{1}). 
Note that in the case $N=1$ (\ref{2.3}) reads
\begin{equation}\label{2.4} 
f(\mu) = {1 \over 2 \pi} \int_0^{2 \pi} 
{(1 - |\mu|^2) \over |1 - \mu^* e^{i \theta} |^2}f( e^{i \theta}) \, 
d\theta,
\end{equation}
which is the celebrated Poisson formula on a circle, giving the value of
an analytic function $f$ for $|\mu|  < 1$ in terms of its value on the unit
circle.  Thus (\ref{2.3}) can be regarded as an $N$-dimensional generalization
of this result.

The multivariable orthogonal polynomials corresponding to (\ref{1})
 are the symmetric Jack polynomials
$P_\kappa^{(2/\beta)}(z)$ \cite{St89,Ma95}, as they possess the orthogonality
property
\begin{equation}\label{2.5}
\prod_{l=1}^N \int_0^{2 \pi} d \theta_l \, P_\kappa^{(2/\beta)}(z)
P_\sigma^{(2/\beta)}(z^*)
 \prod_{1 \le j < k \le N} | e^{i \theta_k} -
e^{i \theta_j}|^\beta = {\cal N}_\kappa \delta_{\kappa,\sigma},
\end{equation}
and form a complete set for the space of analytic functions.
 Here the labels
$\kappa$ and $\sigma$ are  partitions consisting of $N$ parts or less,
and $z:=(e^{i\theta_1},\dots,e^{i\theta_N})$. The normalization
${\cal N}_\kappa$ is given by
\begin{equation}\label{2.6}
{{\cal N}_\kappa \over {\cal N}_0} =
{ P_\kappa^{(2/\beta)}(1^N) d_\kappa' \over
[1 + \beta (N-1)/2]_\kappa^{(2/\beta)} }
\end{equation}
where 
$$
d_\kappa' := \prod_{s \in \kappa} \Big ( {2 \over \beta} a(s) +
l(s) + 1 \Big ), \quad [u]_\kappa^{(\alpha)} :=
\prod_{j=1}^N {\Gamma(u - {\beta \over 2}(j-1) + \kappa_j) \over
\Gamma(u - {\beta \over 2}(j-1)}.
$$
The notation $s \in \kappa$ refers to the diagram of the partition $\kappa$,
and $a(s) = \kappa_i - j$ is the arm length while
$l(s) = \kappa_j' - i$ is the leg length ($\kappa'$ refers to the conjugate
partition of $\kappa$); see e.g.~\cite{Ma95}. The normalization
${\cal N}_0$ is the same quantity as the  normalization
$C_{\beta N}$ in (\ref{1}), and has the explicit value
\begin{equation}\label{2.7}
{\cal N}_0 = C_{\beta N} = (2 \pi)^N
{(N \beta/2)! \over (\beta / 2)!^N}.
\end{equation}
The quantity $ P_\kappa^{(2/\beta)}(1^N)$ in (\ref{2.6})
also has an explicit evaluation
\cite{St89,Ma95,BF97c}, but it suits our purposes to leave it unevaluated.

With these preliminaries, we now pose the problem of specifying the
kernel $K(w,z^*)$, $w = (e^{i\phi_1},$$\dots,e^{i\phi_N})$, Im$(\phi_j) >0$,
 which is an
analytic function of $z^*$, and has the reproducing property
\begin{equation}\label{2.8}
f(w)  =  \prod_{l=1}^N \int_0^{2 \pi} d \theta_l \, K(w,z^*) f(z)
 \prod_{1 \le j < k \le N} | e^{i \theta_k} -
e^{i \theta_j}|^\beta,
\end{equation}
for $f$ analytic and symmetric in $w_1,\dots,w_N$. 
The proof of (\ref{2.3}) will then consist first of evaluating
$K(w,z^*)$ at $w = (\mu,\dots,\mu)$, then transforming the
equation (\ref{2.8}) so that $K(w,z^*)$ becomes real.
 Using the
orthogonality property (\ref{2.5}), as well as the completeness of
$\{ P_\kappa^{(2/\beta)}(z) \}$ for analytic functions, it is a simple
exercise to check that the required kernel $K$ is uniquely given by
\begin{eqnarray}\label{2.9}
K(w,z^*) &  = &\sum_\kappa {P_\kappa^{(2/\beta)}(w)
 P_\kappa^{(2/\beta)}(z^*) \over {\cal N}_\kappa } \nonumber \\
& = &  {1 \over {\cal N}_0} \sum_\kappa 
[1 + \beta (N-1)/2]_\kappa^{(2/\beta)} {P_\kappa^{(2/\beta)}(w)
P_\kappa^{(2/\beta)}(z^*) \over d_\kappa' P_\kappa^{(2/\beta)}(1^N)}.
\end{eqnarray}
where the second equality follows from (\ref{2.6}).
But in general the generalized hypergeometric function
${}_1^{}{\cal F}_{0}^{(2/\beta)}(a;w,z^*)$ is defined by
\begin{equation}\label{2.10}
{}_1^{}{\cal F}_{0}^{(2/\beta)}(a;w,z^*)
= \sum_\kappa
[a]_\kappa^{(2/\beta)} 
{P_\kappa^{(2/\beta)}(w)
P_\kappa^{(2/\beta)}(z^*) \over d_\kappa' P_\kappa^{(2/\beta)}(1^N)}
\end{equation}
so we have
\begin{equation}\label{2.11}
K(w,z^*) = {1 \over {\cal N}_0} {}_1^{}{\cal F}_{0}^{(2/\beta)}
(1 + \beta (N-1)/2;w;z^*).
\end{equation}

In general there is no known expression for 
${}_1^{}{\cal F}_{0}^{(2/\beta)}(a;w,z^*)$ in terms of elementary
functions. However, the case $w = (\mu,\dots,\mu)$ is an exception,
for then we have
\begin{equation}\label{2.11.1}
{}_1^{}{\cal F}_{0}^{(2/\beta)}(a;w,z^*) \Big |_{w = (\mu,\dots,\mu)}
= {}_1^{}{\cal F}_{0}^{(2/\beta)}(a;\mu z^*) :=
\sum_\kappa {\mu^{|\kappa|} [a]_\kappa^{(2/\beta)} P_\kappa^{(2/\beta)}(z^*)
\over d_\kappa'}.
\end{equation}
The significance of this is that ${}_1^{}{\cal F}_{0}^{(2/\beta)}(a;\mu z^*)$
can be summed according the generalized binomial formula
\cite{Ka93}
\begin{equation}\label{2.12}
{}_1^{}{\cal F}_{0}^{(2/\beta)}(a;\mu z^*)
= \prod_{j=1}^N {1 \over (1 - \mu z_j^*)^a }, \quad |\mu z_j| < 1.
\end{equation}
Comparing (\ref{2.11}) -- (\ref{2.12}) we therefore have
\begin{equation}\label{2.13}
K(w,z^*) \Big |_{w = (\mu,\dots,\mu)} = {1 \over {\cal N}_0}
\prod_{j=1}^N {1 \over (1 - \mu z_j^*)^{1 + \beta (N-1)/2}}.
\end{equation}
Although this is an explicit solution to the problem of determining the
kernel in (\ref{2.8}) in the case $w = (\mu,\dots,\mu)$, it does
not immediately establish (\ref{2.3}) as the kernel (\ref{2.13}) is
not real. Note that in the case $N=1$, (\ref{2.13}) is the Cauchy kernel
from elementary complex analysis. For general $N$, to obtain a real
(Poisson) kernel from the Cauchy kernel, we proceed as in the one
dimensional case and simply make the replacement
$$
f \mapsto \prod_{j=1}^N {1 \over (1 - \mu^* z_j)^{1 + \beta (N - 1)/2}} f
$$
in (\ref{2.8}) with $w = (\mu,\dots,\mu)$. The formula (\ref{2.3}) results,
thus establishing its validity for general $\beta > 0$.

\subsection{Fluctuation formulas}
In the application of random matrix theory, an important class of 
observables are the linear statistics $A = \sum_{j=1}^N a(\lambda_j)$.
The first two moments of these statistics are given by
\begin{eqnarray}
\langle A \rangle & = & \int_0^{2 \pi}
 \rho_{(1)}(\theta) a(\theta) \, d\theta \label{e.00}\\
{\rm Var} (A) & := &
\int_0^{2 \pi}  d\theta_1 \, a(\theta_1) \int_0^{2 \pi}  d\theta_2 \, a(\theta_2)
S(\theta_1,\theta_2) \label{e.0}
\end{eqnarray}
where $S(\lambda_1,\lambda_2)$ denotes the structure function
$$
S(\lambda_1,\lambda_2) := \rho_{(2)}^T(\lambda_1,\lambda_2) +
\rho_{(1)}(\lambda_1) \delta(\lambda_1 - \lambda_2),
$$
with $\rho_{(1)}$ denoting the density and $\rho_{(2)}^T$ denoting the
truncated two particle distribution function.
In this subsection these quantities will be considered for the eigenvalue
p.d.f.~(\ref{6}).

It is instructive to first consider the case $\mu = 0$, when
(\ref{6}) reduces to the eigenvalue 
p.d.f.~(\ref{1}) for the circular ensemble. In this case \cite{Fo95d}
\begin{eqnarray}
 \rho_{(1)}(\theta^{(0)}) & = & {N \over 2 \pi} \label{e.10}\\
S(\theta^{(0)}_1, \theta^{(0)}_2) & \mathop{\sim}\limits_{N \to \infty} &
- {1 \over \beta \pi^2}
{\partial^2 \over \partial \theta^{(0)}_1 \partial \theta^{(0)}_2}
\log | \sin ( \theta^{(0)}_1 -  \theta^{(0)}_2)/2 | \label{e.1}
\end{eqnarray}
(the use of the superscript $(0)$ indicates that $\mu = 0$ in (\ref{6})),
where in the asymptotic expression (\ref{e.1}) all oscillatory terms,
each of which have period some integer multiple of $2 \pi /N$,
are ignored. For example, we have the exact result (see e.g.~\cite{Me91})
\begin{eqnarray*}
\rho_{(2)}^T(\theta^{(0)}_1,\theta^{(0)}_2)  & = &
- \Big ( {1 \over 2 \pi} \Big )^2 {\sin^2 N 
( \theta^{(0)}_1 -  \theta^{(0)}_2)/2 \over
\sin^2( \theta^{(0)}_1 -  \theta^{(0)}_2)/2} \\ & = &
- {1 \over 2} \bigg ( {1 \over 2 \pi} \Big )^2
\Big ( {1 \over \sin^2( \theta^{(0)}_1 -  \theta^{(0)}_2)/2} -
{\cos N ( \theta^{(0)}_1 -  \theta^{(0)}_2) \over 
\sin^2( \theta^{(0)}_1 -  \theta^{(0)}_2)/2} \bigg ).
\end{eqnarray*}
Ignoring the oscillatory term with the factor of 
$\cos N ( \theta^{(0)}_1 -  \theta^{(0)}_2)$ gives
$$
\rho_{(2)}^T(\theta^{(0)}_1,\theta^{(0)}_2) 
\mathop{\sim}\limits_{N \to \infty} 
- {1 \over 2 (2 \pi)^2} {1 \over \sin^2( \theta^{(0)}_1 -  \theta^{(0)}_2)/2}
\: = \: - {1 \over 2 \pi^2} 
{\partial^2 \over \partial \theta^{(0)}_1 \partial \theta^{(0)}_2}
\log | \sin ( \theta^{(0)}_1 -  \theta^{(0)}_2)/2 |
$$
vaild for $ \theta^{(0)}_1 \ne  \theta^{(0)}_2$. The validity of
(\ref{e.1}) for $ \theta^{(0)}_1 =  \theta^{(0)}_2$ is then deduced
by noting that if we write (define)
\begin{equation}
\int_0^{2 \pi} 
{\partial^2 \over \partial \theta^{(0)}_1 \partial \theta^{(0)}_2}
\log | \sin ( \theta^{(0)}_1 -  \theta^{(0)}_2)/2 | \, d\theta_1^{(0)} =
{\partial \over \partial \theta_2^{(0)}} 
\int_0^{2 \pi} 
{\partial \over \partial \theta^{(0)}_1 }
\log | \sin ( \theta^{(0)}_1 -  \theta^{(0)}_2)/2 | \, d\theta_1^{(0)}
\label{pers}
\end{equation}
then this integral vanishes. This is the perfect screening sum rule
for the underlying log-gas system, and is a fundamental requirement of
$S(\theta^{(0)}_1, \theta^{(0)}_2)$ \cite{Ma88}.

Substituting (\ref{e.1}) in (\ref{e.0}), interchanging the order of
differentiation and integration according to (\ref{pers}), and using the
Fourier expansion
$$
\log | \sin ( \theta^{(0)}_1 -  \theta^{(0)}_2)/2 | =
\sum_{p=-\infty}^\infty \alpha_p e^{i p ( \theta^{(0)}_1 -  \theta^{(0)}_2)},
\qquad \alpha_p = - {1 \over 2 |p|} \: \: (p \ne 0), \: \alpha_0 =
-2 \pi \log 2,
$$
allows (\ref{e.0}) to be evaluated as
\begin{equation}\label{f.1a}
\mbox{Var}(A)^{(0)} = {4 \over \beta} \sum_{n=1}^\infty n a_n a_{-n},
\qquad a(\theta) = \sum_{n=-\infty}^\infty a_n e^{i n \theta}.
\end{equation}
This has the well known feature of being independent of $N$ -- fluctuations
are therefore strongly suppressed. Furthermore, it has been rigorously
proved by Johansson \cite{Jo88,Jo95} that the corresponding
full distribution of the linear
statistic $A$ is given by the central limit-type theorem
\begin{equation}\label{f.2a}
{\rm Pr}(u=A)  \mathop{\sim}\limits_{N \to \infty}
{1 \over (2 \pi \sigma^2)^{1/2}}
e^{-(u - \langle A \rangle)^2/2 \sigma^2}
\end{equation}
where $\langle A \rangle$ is given by (\ref{e.00}) with the substitution
(\ref{e.10}), and $\sigma^2 = {\rm Var}(A)$ as given by (\ref{f.1a}).

Let us now consider the situation for general $\mu$, $|\mu| < 1$. 
It is well known
\cite{Hu63}, and is simple to verify, that under the transformation
\begin{equation}\label{f.3a}
e^{i \theta_j} = {e^{i \theta_j^{(0)}} - \mu \over 1 - \mu^* 
e^{i \theta_j^{(0)}}}
\end{equation}
(note that the RHS has unit modulus and this mapping is one-to-one), the
p.d.f.~for general $\mu$ is transformed into the p.d.f.~with $\mu = 0$
according to
\begin{eqnarray}
\prod_{j=1}^N \bigg (
{1 - |\mu|^2 \over | 1 - \mu^* e^{i \theta_j}|^2} \bigg )^{\beta (N-1)/2 + 1}
\prod_{1 \le j < k \le N} | e^{i \theta_k} -
e^{i \theta_j}|^\beta d\theta_1 \cdots d\theta_N \nonumber \\
= 
\prod_{1 \le j < k \le N} | e^{i \theta_k^{(0)}} -
e^{i \theta_j^{(0)}}|^\beta d\theta_1^{(0)} \cdots d\theta_N^{(0)}. \label{f.4}
\end{eqnarray}
This means that the correlation functions for general $\mu$ can be obtained
from the correlation functions for $\mu=0$ by applying the inverse of the
transformation (\ref{f.3a}),
\begin{equation}\label{f.5}
e^{i \theta_j^{(0)}} = {\mu + e^{i \theta_j} \over
1 + \mu^* e^{i \theta_j}},
\end{equation}
and noting that
\begin{equation}\label{f.6}
d \theta^{(0)} = { (1 - |\mu|^2) \over |1 - \mu^* e^{i \theta} |^2 }
d \theta.
\end{equation}
Hence, from (\ref{e.10}), we have
\begin{equation}\label{f.7}
\rho_{(1)}(\theta) = {N \over 2 \pi}
{(1 - |\mu|^2) \over |1 - \mu^* e^{i \theta} |^2 }
\end{equation}
independent of $\beta$, and so from (\ref{e.00}),
\begin{equation}\label{f.8}
\langle A \rangle = {N \over 2 \pi} \int_0^{2 \pi}
{(1 - |\mu|^2) \over |1 - \mu^* e^{i \theta} |^2 } \rho(\theta) \, d \theta.
\end{equation}

For the structure function, substituting (\ref{f.5}) and (\ref{f.6})
in (\ref{e.1}) gives
\begin{eqnarray}
S(\theta_1^{(0)}, \theta_2^{(0)}) d \theta_1^{(0)} d \theta_2^{(0)}
& := & - {1 \over \beta \pi^2}
\bigg ( {\partial^2 \over \partial \theta_1^{(0)} \partial \theta_2^{(0)} }
\log |e^{i \theta_1^{(0)}} - e^{i \theta_2^{(0)}}|
\bigg )  d \theta_1^{(0)} d \theta_2^{(0)} \nonumber \\
& = & - {1 \over \beta \pi^2} 
\bigg ( {\partial^2 \over \partial \theta_1 \partial \theta_2 }
\log \Big | {e^{i \theta_1} - e^{i \theta_2} \over
|1 + \mu^*e^{i \theta_1}| |1 + \mu^*e^{i \theta_2}|} \Big | \bigg )
d \theta_1 d \theta_2 \nonumber \\
& = & - {1 \over \beta \pi^2} 
\bigg ( {\partial^2 \over \partial \theta_1 \partial \theta_2 }
\log \Big | e^{i \theta_1} - e^{i \theta_2}  \Big | \bigg )
d \theta_1 d \theta_2.
\end{eqnarray}
Thus $S(\theta_1, \theta_2)$ for general $\mu$ is identical to
$S(\theta_1^{(0)}, \theta_2^{(0)}) $ for $\mu=0$, and consequently
\begin{equation}\label{ex}
\mbox{Var}(A)^{(0)} = \mbox{Var}(A).
\end{equation}
This fact illustrates a universality feature of the underlying log-gas:
Var($A$) is invariant with respect to the particular one body potential
modifying (\ref{1}), provided the corresponding one body density is a well
behaved function.

Finally, we note that (\ref{ex}) can be demonstrated via a numerical
experiment. The experiment is performed by first generating random
unitary matrices with uniform distribution (matrices from
the CUE). This can be done by
diagonalizing matrices from the GUE (random Hermitian matrices): the matrix
of eigenvectors, when multiplied by a diagonal matrix with entries
$e^{i \theta_j}$, $(j=1,\dots,n)$ where $\theta_j$ is a random angle between
0 and $2 \pi$ with uniform distribution, gives a matrix belonging to the
CUE. Next we calculate the eigenvalues of each 
matrix (which will have distribution (\ref{1}) with $\beta = 2$), and transform
them according to (\ref{f.3a}) with a specific value of $\mu$. The
resulting eigenvalues will have distribution as on the l.h.s.~of (\ref{f.4}).
For each set $k$ of eigenvalues $\{e^{i \theta_j}\}_{j=1,\dots,N}$ we then
calculate $A_k := \sum_{j=1}^N a(\theta_j)$ for some particular choices of
$a$. From the resulting list of values $\{A_k\}$, the empirical mean and 
standard derivation are calculated according to the usual formulas. In
table 1 we present the result of performing this numerical experiment
with $N$, the dimension of the unitary matrix, equal to 15, and
$a(\theta) = \cos j \theta$ $(j=1,\dots,5)$. These empirical values are
compared with the theoretical prediction for the variance in the limit
$N \to \infty$ as given by (\ref{f.1a}) (note that with
$a(\theta)=\cos j \theta$, $a_j = a_{-j}=1/2$, $a_n=0$ otherwise; thus
(\ref{f.1a}) gives Var$(A) = j/2$).

 \vspace{1cm}

\begin{tabular}{|c|c|c|} \hline
 & Empirical & Theoretical\\ 
$a(\theta)$ & Var$(A)$ & Var$(A)$\\ \hline \hline
$\cos \theta$& 0.509  &0.5\\ \hline
$\cos 2 \theta$ & 0.972 &1\\ \hline
$\cos 3\theta$ & 1.6 &1.5\\ \hline
$\cos 4\theta$ & 1.8 &2\\ \hline
$\cos 5\theta$ & 2.6 &2.5\\ \hline
\end{tabular}

\vspace{.5cm}
\noindent 
{\bf Table 1. } The second column contains the
 empirical variance of the quantity
$\sum_{j=1}^N a(\theta_j)$, with $a(\theta)$ as specified. This was calculated for
500 $15 \times 15$ matrices with  eigenvalue distribution given by the
l.h.s.~of (\ref{f.4}) with $\beta=2$ and $\mu=.5$. The final column contains
the theoretical variance for the same quantity in the $N \to \infty$ limit.

\vspace{1cm}

\section{Laguerre ensemble with an $N$-dependent exponent}
\setcounter{equation}{0}
\subsection{Motivation}
Recently Brouwer et al.~\cite{BFB96} have considered the problem of the
distribution of the eigenvalues of the Wigner-Smith matrix
$ Q = - i \hbar S^{-1} \partial S / \partial E $. Here $S$ refers to the
scattering matrix coupled to a perfect lead which supports $N$ channels of
the same energy $E$. It was found that for each of the three possible
symmetries of $S$, orthogonal ($\beta = 1$), unitary ($\beta = 2$)
and symplectic ($\beta = 4$), the p.d.f.~for the reciprocal of the
eigenvalues of $Q$ is given by (\ref{2}) with $a=N$. This motivates a
study of some of the properties of the distribution functions and
fluctuation formulas associated with (\ref{2}) for general
$a = YN$, $Y >0$.

\subsection{$\beta=1$ and $2$}
As remarked in the introduction, the p.d.f.~(\ref{2}) for $\beta = 1,2$
and 4 is realized as the eigenvalue p.d.f.~of random Wishart matrices
$ A =  X^\dagger  X$, where $ X$ has dimension $M \times
N$. For $\beta = 1$ and 2, and with $a$ proportional to $N$, the
limiting form of the global density and the statistical properties of
the largest and smallest eigenvalues have been extensively studied
(see \cite{Ed88} and references therein). In particular, with
\begin{equation}\label{f.0}
a = YN
\end{equation}
it is known
that the global eigenvalue density is given by
\begin{equation}\label{f.1}
\lim_{N \to \infty} \rho(4Nx) =
\left \{ \begin{array}{ll}
{1 \over \pi x} \sqrt{(x - t_1(Y))(t_2(Y) - x)}, & t_1(Y) < x < t_2(Y) \\
0, & {\rm otherwise} \end{array} \right.
\end{equation}
where
\begin{equation}\label{f.2}
t_1(Y) = {1 \over 4} (\sqrt{1 + Y} - 1)^2, \quad
t_2(Y) = {1 \over 4} (\sqrt{1 + Y} + 1)^2.
\end{equation}

In qualitative terms, the result (\ref{f.1}) says that the support of the
density, which is $(0,1)$ when $Y=0$, is repelled from the origin and
elongated as $Y$ increases. This is consistent with the log-gas interpretation
of (\ref{2}) with $a$ given by (\ref{f.0}), as then there is an external
 charge of 
strength $YN$ placed at the origin. This charge repels the $N$ mobile
charges of unit strength away from the origin. The fact that (\ref{f.1})
is independent of $\beta$ is also consistent with the log-gas interpretation.
In fact macroscopic electrostatics says that the one-body potential in
(\ref{4}) results from a neutralizing background charge density
$\rho_b(y)$ according to
\begin{equation}\label{f.3}
{x \over 2} - {YN \over 2} \log x + C =
\int_I \rho_b(y) \log |x - y| \, dy, \quad x \in I
\end{equation}
where $I$ is an interval in ${\bf R}^+$. The quantity $\beta$
does not appear in this equation, so $\rho_b(y)$ is
independent of $\beta$. But to leading order the particle
density will equal the background density, as in general Coulomb systems
strongly suppress charge fluctuations \cite{Ma88}. The expected
independence of (\ref{f.1}) on $\beta$ follows.

\subsection{$\beta$ even}
For $\beta$ even, an exact $\beta$-dimensional integral representation of the
density in the finite system is available \cite{Fo94b,BF97}, which allows
 the global density limit to be taken explicitly. 

Now, in a system of $N+1$ particles, the one-body density $\rho(x)$ in the
Laguerre ensemble is given by
\begin{equation}\label{f.44}
\rho_{N+1}(x) := {N+1 \over Z_{N+1}(a,\beta)} e^{-\beta x/2} 
x^{\beta a/2} I_N(a,\beta;x)
\end{equation}
where
\begin{eqnarray}
 Z_{N+1}(a,\beta) & := & \prod_{l=1}^{N+1} \int_0^\infty
dx_l \, e^{-\beta x_l/2} x_l^{\beta a/2} \prod_{1 \le j < k \le
N+1} |x_k - x_j|^{\beta} \label{Z}\\
I_N(a,\beta;x)  & := & \prod_{l=1}^{N} \int_0^\infty
dx_l \, |x - x_l|^\beta 
e^{-\beta x_l/2} x_l^{\beta a/2} \prod_{1 \le j < k \le
N} |x_k - x_j|^{\beta} \label{I}
\end{eqnarray}
The normalization (\ref{Z}) is a well known limiting case of the Selberg
integral, and has the exact evaluation (see e.g.~\cite{As80})
\begin{equation}\label{ZE}
 Z_{N}(a,\beta) =
\left ( {2 \over \beta} \right )^{N(1 + \beta a/2) + \beta N (N-1)/2}
\prod_{j=0}^{N-1} { \Gamma(1+\beta(j+1)/2) \Gamma(1+\beta(j+a)/2) \over
\Gamma(1 + \beta/2)}.
\end{equation}

Notice that for $\beta$ even the integral (\ref{I}) is a polynomial in $x$.
In this case  (\ref{I}) has been shown to be expressible in terms of a
certain generalized Laguerre polynomial based on Jack polynomials
\cite{Fo94b}. Furthermore, this generalized Laguerre polynomial has a
different integral representation, which allows the
$N$-dimensional integral (\ref{I}) to be expressed as a
$\beta$-dimensional integral. This reads \cite{BF97}
\begin{equation}\label{s.1}
{1 \over Z_N(a+2,\beta)} I_N(a,\beta;x) =
{1 \over Q(a,\beta)} | f(a,\beta;x) |
\end{equation}
where
\begin{eqnarray}\label{s.2}
\lefteqn{f(a,\beta;x) := } \nonumber \\&&
\int_{{\cal C}_1} dt_1 \cdots \int_{{\cal C}_\beta} dt_\beta 
\prod_{j=1}^\beta e^{xt_j} t_j^{-N-3+2/\beta} 
(1 - t_j)^{a+N+2/\beta - 1} \prod_{1 \le j < k \le \beta} (t_k - t_j)^{4/\beta}.
\end{eqnarray}
The contours of integration in (\ref{s.2}) must be simple loops which start
at $x = 1$ and enclose
the origin, and the quantity $Q(a,\beta)$ in (\ref{s.1}) is chosen so that
at $x=0$, the r.h.s.~equals 1 (the l.h.s.~has this property). 
Also, the modulus sign in (\ref{s.1}) has been included for convenience
to eliminate terms of unit modulus which otherwise occur in
$Q(a,\beta)$; this is valid for $x \in {\bf R}$ since then the l.h.s.~is
positive. Choosing 
each ${\cal C}_j$ in (\ref{s.2}) to be the unit circle gives
\begin{equation}\label{s.3}
Q(a,\beta) = (2 \pi)^\beta M_\beta(a+2/\beta-1,N,2/\beta)
\end{equation}
where
\begin{eqnarray}\label{s.4}
M_N(a',b',c') & := & \prod_{l=1}^N \int_{-1/2}^{1/2}
e^{\pi i \theta_l (a' - b')} | 1 + e^{2 \pi i \theta_l}|^{a' + b'}
\prod_{1 \le j < k \le N} |  e^{2 \pi i \theta_k} -
 e^{2 \pi i \theta_j} |^{2c'} \nonumber \\
& = & \prod_{j=1}^N{\Gamma(a'+b'+1 + (j-1)c') \Gamma(1+c'j) \over
\Gamma(a'+1 + (j-1)c') \Gamma(b'+1 + (j-1)c') \Gamma(1+c')}
\end{eqnarray}
(the integral $M_N$ is due to Morris \cite{Mo82}).

Our interest is in the large-$N$ asymptotic form of $\rho_{N+1}(
4Nx) \Big |_{a = YN}$, which from (\ref{f.44}) and (\ref{s.1}) requires the
large-$N$ form of $f(YN,\beta;x)$. The necessary technique is a
generalized saddle point analysis as introduced in \cite{Fo94b} and presented
in detail in \cite{BF97}. The saddle points occur at the stationary points
of
\begin{equation}\label{s.5a}
N \Big ( 4xt_j - \log t_j + (Y + 1) \log (1 - t_j) \Big ),
\end{equation}
which is the $N$-dependent term of the integrand of
 $f(YN,\beta;x)$ when expressed as an exponential. A simple calculation
gives that there are two stationary points, $t_+$ and $t_-$ say, given by
\begin{equation}\label{s.5}
t_{\pm} = {x - Y/4 \pm ((x - Y/4)^2 - x)^{1/2} \over 2x}.
\end{equation}
Note that $t_+ = t_-^*$ for
\begin{equation}\label{s.6}
(x - Y/4)^2 - x < 0.
\end{equation}

Assuming (\ref{s.6}), following the strategy of \cite{BF97}, the leading
large-$N$ asymptotic behaviour is obtained by deforming $\beta / 2$
of the contours through $t_+$, and the remaining $\beta / 2$ of the
contours through $t_-$ (this introduces a factor of $\left ( {\beta 
\atop \beta/2} \right )$ to account for the number of ways of dividing the
$\beta$ contours into these two classes). The calculation now proceeds in
a conventional way, with the exponent (\ref{s.5a}) being expanded 
about $t_{\pm}$ to second order, and the $N$-independent terms in the integrand
replaced by their value at $t_+$ or $t_-$ as appropriate. After this
step we have
\begin{eqnarray}\label{s.7}
|f(YN,\beta;x)| & \sim & \left ( {\beta \atop \beta/2} \right )
| g_2(t_+,Y) |^\beta \Big | \int_{-\infty}^\infty dt_1 \cdots
\int_{-\infty}^\infty dt_{\beta/2}
e^{-{N \over 2} g_1(t_+,Y)(t_1^2 + \cdots + t_{\beta/2}^2)} \nonumber \\
&& \hspace*{3cm} \times
\prod_{1 \le j < k \le \beta / 2} |t_k - t_j|^{4/\beta} \Big |^2 
\nonumber \\
& = & \left ( {\beta \atop \beta/2} \right ) | g_2(t_+,Y) |^\beta
| N g_1(t_+,Y)|^{-\beta + 1} \Big ( V_{\beta / 2} (2/\beta) \Big )^2
\end{eqnarray}
where
\begin{eqnarray}
g_1(t_+,Y) & := & {1 \over t_+^2} - {1 + Y \over (1 - t_+)^2}\label{g.1}\\
g_2(t_+,Y) & := & e^{N(4xt_+ - \log t_+ + (Y + 1) \log (1 - t_+))}
t_+^{-3+2/\beta}(1 - t_+)^{2/\beta - 1} (t_+ - t_-) \label{g.2}\\
V_N(c) & := & \int_{-\infty}^\infty dt_1 \cdots \int_{-\infty}^\infty dt_N
e^{- { 1 \over 2}(t_1^2 + \cdots + t_N^2)} 
\prod_{1 \le j < k \le N} |t_k - t_j|^{2c} \nonumber \\
& = & (2 \pi)^{N/2} \prod_{j=0}^{N-1} {\Gamma(1 + c(j + 1)) \over
\Gamma(1 + c)}. \label{s.8}
\end{eqnarray}
The equality in (\ref{s.7}) follow from a simple change of variables,
while (\ref{s.8}) is known as Mehta's integral, and can be evaluated
as a limiting case of the Selberg integral \cite{As80} (for a direct
evaluation see \cite{Ev94}).

{}From the definition (\ref{s.5}) we have that
$$
|x_+|^2 = {1 \over 4x}, \quad |1 - x_+|^2 = {1 + Y \over 4x},
\quad |t_+ - t_-|^2 = {1 \over x^2} \Big ( x - (x - Y/4)^2 \Big ),
$$ 
and use of these results in the formulas (\ref{g.1}) and (\ref{g.2})
defining $g_1$ and $g_2$ shows
\begin{equation}\label{s.10a}
{|g_2(t_+,Y)|^{\beta} \over | g_1(t_+,Y)|^{\beta - 1}} =
e^{2 \beta N ( x - Y/4)} (1 + Y)^{(1+Y)N \beta / 2} {(4x)}^{-YN \beta / 2}
(1 + Y)^{1/2} {1 \over x} \Big ( x - (x - Y/4)^2 \Big )^{1/2}.
\end{equation}
Furthermore, from (\ref{ZE})
\begin{equation}\label{s.10}
{Z_N(a + 2,\beta) \over Z_{N+1}(a,\beta)} =
\Big ( {2 \over \beta} \Big )^{N\beta - {\beta \over 2}}
{\Gamma(1 + {\beta \over 2}) \over \Gamma(1 + {\beta \over 2} (N + 1))}
{\Gamma({\beta \over 2}(a + N + 1) + 1) \over
\Gamma({\beta \over 2}a + 1) \Gamma({\beta \over 2}(a + 1)+1 )},
\end{equation}
while use of (\ref{s.4}) in (\ref{s.3}) and comparison with (\ref{s.8})
shows that
\begin{equation}\label{s.11}
{1 \over Q(a,\beta)} = {1 \over (2 \pi)^{\beta/2} V_\beta(2/\beta)}
\prod_{j=1}^\beta {\Gamma(a + {2 \over \beta}j)
\Gamma(N + 1 + {2 \over \beta} (j-1)) \over
\Gamma( a + N + {2 \over \beta}j)},
\end{equation}
and use of (\ref{s.8})  gives
\begin{equation}\label{s.12}
{(V_{\beta/2}(2/\beta))^2 \over V_\beta(2/\beta)} = (\beta/2)^{\beta/2}
{\Gamma(1 + {\beta \over 2}) \over \Gamma(1 + \beta)}.
\end{equation}

With (\ref{s.1}) substituted in (\ref{f.44}), 
the remaining task is to use Stirling's formula to compute the
leading large-$N$ asymptotic behaviour of (\ref{s.10}) and (\ref{s.11})
with $a = YN$.
Performing this task, and substituting the resulting expression
together with (\ref{s.10a}) and (\ref{s.12}) in
(\ref{s.1}), shows that for $x$ such that (\ref{s.6}) is true
\begin{equation}\label{s.13}
\lim_{N \to \infty} \rho_{N+1}(4Nx) \Big |_{a = YN} = {1 \over 2 \pi x}
 \Big ( x - (x - Y/4)^2 \Big )^{1/2}.
\end{equation}
Outside the interval (\ref{s.6}), i.e.~outside
$x \in [{1 \over 2}(1+{Y \over 2} -
\sqrt{1+Y}), {1 \over 2}(1+{Y \over 2} + \sqrt{1+Y})]$,
 this limit must vanish. This is seen
from the fact that  the
density is positive, and must satisfy the normalization
$$
\int_0^\infty \rho_{N+1}(4Nx) \, dx \: \sim \: {1 \over 4},
$$
which is satisfied by the r.h.s.~of (\ref{s.13}). The result (\ref{s.13}) for
the scaled density, established for even $\beta$, is identical to the
result (\ref{f.1}) known for $\beta = 1$ and 2, as expected.

\section{The distribution functions in the neighbourhood of the smallest
eigenvalue}
\setcounter{equation}{0}
\subsection{The $n$-point distributions}
For fixed $N$ and $\beta = 1,2$ or 4, the exact expressions for the general
$n$-point distribution function in the Laguerre ensemble
are known \cite{Br65,NW91,NF95}. With $a = YN$ and $N$ large, it is natural to move the
origin to the (mean) location of the smallest eigenvalue, and to scale
the eigenvalues so that the mean spacing near the spectrum edge is 
$O(1)$. One anticipates that the limiting $n$-point distribution function
will correspond to the $n$-point distribution function for the
so-called soft edge, which is the edge of the spectrum for the
Gaussian random matrix ensemble, with the eigenvalues appropriately scaled.

In quantitative terms, we expect that for appropriate $\nu(N)$ 
independent of $\beta$,
\begin{eqnarray}\label{tu.1}
\lim_{N \to \infty}
\Big ( \nu(N) \Big )^n \rho_{(n)}\Big (N(1 - \sqrt{1+Y})^2 - \nu(N)x_1,
\dots, N(1 - \sqrt{1+Y})^2 - \nu(N)x_n \Big ) \nonumber \\ =
\rho_{(n)}^{\rm soft}(x_1,\dots,x_n).
\end{eqnarray}
On the l.h.s., $\rho_{(n)}$ refers to the $n$-point distribution function
for the Laguerre ensemble with $a = YN$. Note from (\ref{f.1})
and (\ref{s.13}) that $N(1 - \sqrt{1+Y})^2$ is the location of the smallest
eigenvalue (to leading order in $N$). On the r.h.s.
\begin{equation}\label{tu.2}
\rho_{(n)}^{\rm soft}(x_1,\dots,x_n) :=
\lim_{N \to \infty}\Big ( {1 \over 2^{1/2} N^{1/6}} \Big )^n
\rho_{(n)}\Big ( (2N)^{1/2} + {x_1 \over 2^{1/2} N^{1/6}}, \dots,
 (2N)^{1/2} + {x_n \over 2^{1/2} N^{1/6}} \Big ),
\end{equation}
where here $\rho_{(n)}$ refers to the $n$-point distribution function
for the Gaussian ensemble defined by the eigenvalue p.d.f
\begin{equation}\label{tu.25}
{1 \over C} \prod_{l=1}^N e^{-\beta x_l^2/2}
\prod_{1 \le j < k \le N} |x_k - x_j|^\beta,
\end{equation}
and $(2N)^{1/2}$ is the leading order location of the largest eigenvalue.
In this section (\ref{tu.1}) will be explicitly verified for $\beta = 2$,
with the quantity $\nu(N)$ shown to be given by
\begin{equation}\label{tu.3}
\nu(N) = N^{1/3} {2^{1/3}(\sqrt{1+Y} - 1)^2 \over
(Y^2 - (2+Y)( \sqrt{1+Y} - 1)^2 )^{1/3}}.
\end{equation}

Once $\nu(N)$ has been determined, the validity of (\ref{tu.1})
for $n=1$ can be established
by matching the one-body density in the neighbourhood
of the smallest eigenvalue implied by
(\ref{s.13}) with the asymptotic behaviour \cite{Fo93a}
\begin{equation}\label{w.1}
\rho_{(1)}^{\rm soft}(x) \mathop{\sim}\limits_{x \to -\infty}
{\sqrt{|x|} \over \pi}
\end{equation}
(this idea is motivated by a similar procedure used in \cite{Ni96,KF97}).
Now the result (\ref{s.13}) implies that for all $\beta$
\begin{equation}\label{w.2}
\rho_{(1)}\Big ( N(1 - \sqrt{1+Y})^2 - \nu(N)x \Big )
\mathop{\sim}\limits_{N \to \infty \atop x \to -\infty} 
{\sqrt{|x|} \over \pi} {(1+Y)^{1/4} \over (1 - \sqrt{1+Y})^2}
\sqrt{\nu(N) \over N},
\end{equation}
valid for $0 \ll x \ll N/\nu(N)$. Substituting (\ref{w.1}) and (\ref{w.2})
in (\ref{tu.1}) with $n=1$, and using (\ref{tu.3}), shows that (\ref{tu.1})
is satisfied provided
\begin{equation}\label{w.3}
(1 + Y)^{1/4} {2^{1/2}(\sqrt{1 + Y} - 1) \over
((2 + Y)(Y^2 - \sqrt{1 + Y} - 1)^2 )^{1/2}}=1,
\end{equation}
which is readily verified.

In preparation for verifying (\ref{tu.1}) for $\beta = 2$ and general $n$, 
we first recall some formulas particular to  that coupling.
For the Laguerre ensemble (\ref{2}) we have \cite{Br65},
\begin{equation}\label{tu.4}
\rho_{(n)}^{\rm soft}(x_1,\dots,x_n) = \det [P_N(x_j,x_k)]_{j,k=1,\dots,n},
\end{equation}
where with $L_n^a(x)$ denoting the Laguerre polynomial of degree $n$,
\begin{eqnarray}\label{tu.5}
P_N(x,y) & := & (xy)^{a/2} e^{-(x+y)/2}  c_N
{L_N^a(x) L_{N-1}^a(y) - L_N^a(y) L_{N-1}^a(x) \over x-y}, \nonumber \\
&:= & (xy)^{a/2} e^{-(x+y)/2}  {c_N \over N + a}
{L_N^a(x) y {L_N^a}'(y) - L_N^a(y) x {L_N^a}'(x) \over x-y},
\: 
c_N = {\Gamma (1+N) \over \Gamma (a+N)}.
\end{eqnarray}
Furthermore, we know that \cite{Fo93a}
\begin{equation}\label{tu.6}
\rho_{(n)}^{\rm soft}(x_1,\dots,x_n) = \det [ K^{\rm soft}(x_j,x_k)
]_{j,k=1,\dots,n}
\end{equation}
where, with ${\rm Ai}(x)$ denoting the Airy function,
\begin{equation}\label{tu.7}
 K^{\rm soft}(x,y) := {{\rm Ai}(x){\rm Ai}'(y) - {\rm Ai}(y){\rm Ai}'(x)
\over x- y}.
\end{equation}
Comparison of (\ref{tu.6}) and (\ref{tu.4}), and the fact that (\ref{tu.1})
is valid for $n=1$ once (\ref{tu.3}) is established 
shows that to verify
(\ref{tu.1}) it suffices to establish the asymptotic formula
\begin{equation}\label{tu.8}
x^{NY/2} e^{-x/2} L_N^{YN}(x) \Big |_{x=N(1 - \sqrt{1+Y})^2 - \nu(N) X}\:
\mathop{\sim}\limits_{N \to \infty} \: k_N(Y) {\rm Ai}(X)
\end{equation}
with   $\nu(N)$ given by (\ref{tu.3}) and $\mu$ fixed. There is no need
to specify $k_N(Y)$, as its value is uniquely determined by (\ref{w.1})
and (\ref{w.2}). Indeed, substituting (\ref{tu.8}) in (\ref{tu.5}), then
substituting the resulting expression in (\ref{tu.4})
with $n=1$, setting $x=y$ and 
comparing with (\ref{tu.1}) for
$x \to -\infty$ shows that
\begin{equation}\label{tu.9}
\Big (k_N(Y) \Big )^2 =  {1 \over
(1 - \sqrt{1 + Y})^2} {\nu(N) \over c_N}.
\end{equation}
In table 2 we give the numerical value of the ratio of the l.h.s.~to r.h.s.~of
(\ref{tu.8}) with $k_N(Y)$ given by (\ref{tu.9}) for various values of
$X,Y,N$.

 \vspace{1cm}

\begin{tabular}{c|ccc} 
 & $(0,1)$ & $(1,2)$ & $(-1,2)$ \\ \hline
50 & 1.0426 & 1.0567 & 0.9929 \\
60 & 1.0402 & 1.0556 & 0.9938 \\
70 & 1.0383 & 1.0544 & 0.9944 \\
80 & 1.0367 & 1.0533 & 0.9949 \\
90 & 1.0353 & 1.0523 & 0.9953
\end{tabular}

\vspace{.5cm}
\noindent 
{\bf Table 2. } Numerical value of the ratio of the  l.h.s.~to r.h.s.~of
(\ref{tu.8}) for various values of $N$ (column on left) and $(X,Y)$ (top
row).

\vspace{1cm}

The asymptotic formula (\ref{tu.8}) can be derived by utilizing the fact
that $y= e^{-x/2} x^{(\alpha + 1)/2} L_N^\alpha (x)$ satisfies the second
order differential equation
\begin{equation}\label{tu.10}
y'' + \Big ( {2N + \alpha + 1 \over 2 x} +
{1 - \alpha^2 \over 4x^2} - {1 \over 4} \Big ) y = 0.
\end{equation}
Substituting $\alpha = YN$, $x = N(1 - \sqrt{1 + Y})^2
- \nu(N) X$, shows that for large $N$ (\ref{tu.10}) reduces to
\begin{equation}\label{tu.11}
y'' - {1 \over 2(1 - \sqrt{1 + Y})^6} 
\Big ( Y^2 - (2 + Y)(1 - \sqrt{1 + Y})^2 \Big )
{(\nu(N))^3 \over N} xy = 0.
\end{equation}
With $\nu(N)$ given by (\ref{tu.3}), this equation reads
$y'' - xy=0$ and its  unique solution 
which decays as $X \to - \infty$ is $y = k_N(Y) {\rm Ai}(X)$, thus
establishing (\ref{tu.8}).

\subsection{Distribution of the smallest eigenvalue}
In principle, knowledge of the $n$-point distributions allows the 
calculation of other statistical quantities such as the distribution of the
smallest eigenvalue. This together with 
the above results implies that, after appropriate change
of origin and scale, the p.d.f.~of the smallest eigenvalue in the Laguerre
ensemble with $a=YN$ equals, in the $N \to \infty$ limit, the p.d.f.~for the
largest (or equivalently smallest) eigenvalue in the Gaussian ensemble
(\ref{tu.25}). For $\beta = 2$, this can be explicitly verified from the
non-linear equations characterizing the respective p.d.f.'s due to Tracy
and Widom \cite{TW94a,TW94b}.

Let $E(x)$ denote the probability that the interval $(0,x)$ in the
Laguerre ensemble (\ref{2}) with $\beta = 2$ contains no eigenvalues, and
let
\begin{equation}\label{sigma}
\sigma(x) = x {d \over dx} \log E(x).
\end{equation}
Then it has been derived in \cite{TW94b} that $\sigma(x)$ satisfies the
non-linear equation
\begin{eqnarray}\label{sigma1}
(x \sigma'')^2 & = & 4x(\sigma')^3 + \sigma^2 + (2a + 4N - 2x) \sigma \sigma'
\nonumber \\&& +(a^2 -2ax - 4Nx + x^2)(\sigma')^2 - 4 \sigma (\sigma')^2.
\end{eqnarray}
Also, let $\tilde{E}(x)$ denote the probability that there are no
eigenvalues between $(x,\infty)$ in the (infinite dimensional)
Gaussian ensemble (\ref{tu.25}) with coordinates as in (\ref{tu.2}). Then it
was derived in \cite{TW94a} that the quantity
\begin{equation}\label{sigma2}
R(x) := {d \over dx} \log \tilde{E}(x)
\end{equation}
satisfies the non-linear equation
\begin{equation}\label{sigma3}
(R'')^2 + 4 R' \Big ( (R')^2 - xR' + R \Big ) = 0.
\end{equation}

As the p.d.f.~for the smallest (largest) eigenvalue is simply related to
$E(x)$ ($\tilde{E}(x)$) by differentiation, we see that the d.e.~(\ref{sigma1})
characterizes the p.d.f.~for the smallest eigenvalue in the Laguerre
ensemble with $\beta = 2$, while (\ref{sigma3}) characterizes the limiting
form of the p.d.f.~of the eigenvalue at the spectrum edge of the
Gaussian ensemble. (Of course boundary conditions must be specified; these
follow from the small (large) $x$ behaviour of the density.)
The results of the previous subsection imply that
\begin{equation}\label{sigma4}
\lim_{N \to \infty} E \Big ( N(\sqrt{1 + Y} - 1)^2 - \nu(N) x \Big )
\Big |_{a = YN} = \tilde{E}(x).
\end{equation}
Indeed, a straightforward calculation using the explicit form 
(\ref{tu.3}) of $\nu(N)$ shows that after changing variables
$x \mapsto N(\sqrt{1 + Y} - 1)^2 - \nu(N) x$ in (\ref{sigma1}) and
introducing the function
\begin{equation}\label{sigma5}
\tilde{\sigma}(x) := {1 \over \nu(N)}
\sigma \Big ( N(\sqrt{1+Y} - 1)^2 - \nu(N) x \Big ),
\end{equation}
the d.e.~(\ref{sigma1}) reduces down to (\ref{sigma3}) with $R = \tilde{\sigma}$.
This is precisely what is required by (\ref{sigma4}).

\section*{Acknowledgement}
This work was supported by the Australian Research Council.


\end{document}